\documentclass[12pt]{article}
\usepackage{amsmath,amssymb,bm,graphicx}
\usepackage{color} 

\begin{document}
\begin{flushright}
\end{flushright}

\makeatletter
\@addtoreset{equation}{section}
\def\theequation{\thesection.\arabic{equation}}
\makeatother


\begin{center}
{\Large{\bf A classical limit of Grover's algorithm induced by dephasing: Coherence vs entanglement 
}}
\end{center}
\vskip .5 truecm
\centerline{\bf  Kazuo Fujikawa$^1$, C.H. Oh$^{2}$  and Koichiro Umetsu$^{3}$
}
\vskip .4 truecm
\centerline {\it $^1$ 
 Interdisciplinary Theoretical and Mathematical Sciences Program,}
\centerline {\it RIKEN, Wako 351-0198, Japan}
\vspace{0.3cm}
\centerline {\it $^2$ Centre for Quantum Technologies and Physics Department,  }
\centerline {\it  National University of Singapore,
Singapore 117543, Singapore}
\vspace{0.3cm}
\centerline {\it $^3$ 
Laboratory of Physics, College of Science and Technology, }
\centerline {\it and Junior College, Funabashi Campus,}
\centerline {\it  Nihon University,  Funabashi, Chiba 274-8501, Japan}
\vspace{0.3cm}

\begin{abstract}
A new approach to the classical limit of Grover's algorithm is discussed by assuming a very rapid  dephasing of a system between consecutive Grover's unitary operations, which drives pure quantum states to decohered mixed states. One can identify a specific element among $N$   unsorted elements by a probability of the order of unity after $k\sim N$ steps of classical amplification, which is realized by a combination of  Grover's unitary operation and rapid dephasing, in contrast to $k\sim \pi \sqrt{N}/4$ steps in quantum mechanical amplification. The
initial two-state system with enormously unbalanced existence probabilities, which is realized by a chosen specific state and a superposition  of all the rest of states among $N$ unsorted states,  is crucial in the present analysis of classical amplification. This analysis illustrates Grover's algorithm in  extremely noisy circumstances. A similar increase from $k\sim \sqrt{N}$ to $k\sim N$ steps due to the loss of quantum coherence takes place in the {\em analog} model of Farhi and Gutmann where the entanglement does not play an obvious role. This supports a view that entanglement is crucial in quantum computation to describe quantum states by a set of qubits, but the actual speedup of the quantum computation is based on quantum coherence.
\end{abstract}


\section{Introduction}
Grover's algorithm
  is fundamental in the study of the basic mechanism of speedups of the quantum computer relative to classical  
digital computers.
The algorithm is defined for a search problem of a specific element  among the  unsorted $N$-number of elements  $\{1,2, ... , N\}$ with $N=2^{n}$.  The classical search problem is first transcribed into a quantum search problem; a specification of the quantum problem is important and we mainly work on  a generic qubit model such as an Ising spin-type model~\cite{barahona}.  
The purpose of the present paper is to examine if a classical limit of Grover's algorithm can be defined by assuming a very rapid dephasing and, if defined, what kind of search steps are required. It will be interesting to see if the quantum $\sim \sqrt{N}$ steps are maintained or the classical $\sim N$ steps are recovered, or something new appears. This analysis is also useful to understand the role of entanglement in quantum search problems, as is explained later. Practically, this analysis illustrates Grover's algorithm in extremely noisy circumstances.

Following the general idea of measurement theory in quantum mechanics~\cite{nielsen}, one may start with an initial state that is a superposition of $N$-number of quantum states with equal probability together with an ancilla $|0\rangle$
\begin{eqnarray}\label{initial-state}
\frac{1}{\sqrt{N}}\sum_{i=1}^{N}|i\rangle\otimes |0\rangle.
\end{eqnarray}
The unitary operator that searches the specific element $a$, which is represented by a quantum state $|a\rangle$, is given by
\begin{eqnarray}\label{measurement-operator}
\tilde{U}=\sum_{i\neq a}|i\rangle\langle i|\otimes I +
|a\rangle\langle a|\otimes X
\end{eqnarray}
where 
\begin{eqnarray}
X=\left(\begin{array}{cc}
            0&1\\
            1&0
            \end{array}\right)
\end{eqnarray}
with $X|0\rangle=|1\rangle$ and $X|1\rangle=|0\rangle$.  The fact that we can identify the state $|a\rangle$ implies that we have some means to distinguish $|a\rangle$ from the rest of states. This identification is also required in the classical search and thus not specific to the present quantum search. It is also important that we assume the existence of such a state $|a\rangle$ when we search  the entire given Hilbert space and we are not giving the existence (or absence) proof of such a state $|a\rangle$. 
After the application of the unitary measurement operator $\tilde{U}$ in \eqref{measurement-operator}, the state \eqref{initial-state} becomes   
\begin{eqnarray}
\frac{1}{\sqrt{N}}\sum_{i\neq a}^{N}|i\rangle\otimes |0\rangle
+\frac{1}{\sqrt{N}}|a \rangle\otimes |1\rangle.
\end{eqnarray}
When one measures the ancilla states, one finds the state 
$ |1\rangle$ with probability $1/N$ and one identifies the state $|a \rangle$, analogously to the use of the particle path (which corresponds to $ |1\rangle$) to identify the spin direction (which corresponds to $|a\rangle$)
in the Stern-Gerlach experiment. This probability is the same as in the classical search problem without a merit of dealing with the superposition of
states
\begin{eqnarray}\label{initial-state2}
|+\rangle\equiv \frac{1}{\sqrt{N}}\sum_{i=1}^{N}|i\rangle. 
\end{eqnarray}
In this paper, we shall first recapitulate the essence of Grover's algorithm which speeds up the data search and then study a classical limit of Grover's algorithm using the idea of dephasing a pure state to a completely mixed state.

\section{Grover's algorithm}
The efficient algorithm of Grover~\cite{grover1,grover2} is based on the measurement operator
\begin{eqnarray}\label{detector}
U=\sum_{i\neq a}|i\rangle\langle i| - |a \rangle\langle a|
\end{eqnarray}
which corresponds to the choice of the ancilla state as
$\frac{1}{\sqrt{2}}[|0\rangle -|1\rangle]$ in \eqref{initial-state}, or 
\begin{eqnarray}
U=\sum_{all\ i}(-1)^{f(i)}|i\rangle\langle i|
\end{eqnarray}
with the oracle $f(i)=0$ for $i\neq a$ and $f(a)=1$.  How this identification of a specific state is done by the oracle (subroutine) is not the issue at this moment, but it could be time consuming. The appearance of the state $|a\rangle$, which we are looking for,  in the unitary operator \eqref{detector} shows that we have a means to identify the state $|a\rangle$ among the states in the unsorted data but it does not imply that we already know the answer to the search problem. 
We also introduce another unitary operator 
\begin{eqnarray}
V=2|+\rangle\langle +| - I,
\end{eqnarray} 
with $I=\sum_{i}|i\rangle\langle i|$.
The basic procedure is to amplify the target state $|a\rangle$ in the initial state \eqref{initial-state2} that is written as 
\begin{eqnarray}\label{initial state}
|+\rangle&=&\sqrt{\frac{N-1}{N}}\frac{1}{\sqrt{N-1}}\sum_{i\neq a}|i\rangle +\frac{1}{\sqrt{N}}|a\rangle
\nonumber\\
&=& \sqrt{\frac{N-1}{N}}|b\rangle +\frac{1}{\sqrt{N}}|a\rangle
\end{eqnarray}
where 
\begin{eqnarray}\label{b-state}
|b\rangle \equiv \frac{1}{\sqrt{N-1}}\sum_{i\neq a}|i\rangle
\end{eqnarray}
with $\langle b|b\rangle=1$ and $\langle a|b\rangle=0$, and then we measure the final state by the basis $\{ |i\rangle \}_{i=1}^{N}$.  Note that the initial state $|+\rangle$ in  \eqref{initial-state2}, which is a superposition of all the states with equal probability,  is written as a superposition of two states $ |a\rangle$ and $|b\rangle$ with extremely unbalanced probability. 
We also define 
\begin{eqnarray}
|-\rangle\equiv \sqrt{\frac{N-1}{N}}|a\rangle -\frac{1}{\sqrt{N}}|b\rangle
\end{eqnarray}
which is normalized and orthogonal to $|+\rangle$,  $\langle -|+\rangle=0$.

When one defines a new complete orthonormal basis set  
using the superposition principle of quantum mechanics 
\begin{eqnarray}\label{new basis}
\{ |a\rangle, |b\rangle, |3^{\prime}\rangle, |4^{\prime}\rangle, ... |N^{\prime}\rangle \},
\end{eqnarray}
we have 
\begin{eqnarray}\label{amplification}
U=|b \rangle\langle b| - |a \rangle\langle a|, \ \ 
V=|+\rangle\langle +| - |-\rangle\langle -|,
\end{eqnarray} 
which are relevant to the present problem of the amplification of the state $|a\rangle$ in the initial state $|+\rangle$ in \eqref{initial state}. It is important that these two unitary operators are defined in the limited 2-dimesional space spanned by  
$\{|a\rangle, |b\rangle\}$ of the entire Hilbert space \eqref{new basis}, which is also equivalent to the space spanned by  $\{|+\rangle, |-\rangle\}$. Namely, the unitary operators $U$ and $V$ act on the states in the two-dimensional subspace of \eqref{new basis}, but all the $n$ quantum qubits with $2^{n}=N$ are influenced by the unitary operation due to the construction of the state $|b\rangle$ in \eqref{b-state}. 

By defining at $t=0$, for example, 
\begin{eqnarray}
\sin\theta=\langle a|+\rangle=\sqrt{1/N}, \ \ \cos\theta=\langle b|+\rangle=\sqrt{(N-1)/N},
\end{eqnarray}
we have $|+\rangle=\cos\theta |b\rangle +\sin\theta|a\rangle$, $|-\rangle=\cos\theta |a\rangle - \sin\theta|b\rangle$ and 
\begin{eqnarray}\label{1.13}
VU|+\rangle
&=&\cos2\theta |+\rangle +\sin2\theta|-\rangle \nonumber\\
&=&\cos3\theta |b\rangle +\sin3\theta|a\rangle
\end{eqnarray}
and by noting 
\begin{eqnarray}\label{1.16}
(VU)^{k}|a\rangle&=&\cos 2k\theta|a\rangle -\sin 2k\theta|b\rangle, \nonumber\\
(VU)^{k}|b\rangle
 &=&\cos2k\theta |b\rangle +\sin2k\theta |a\rangle, 
\end{eqnarray}
we have for  general $k$, assuming $\theta\simeq 1/\sqrt{N}$,
\begin{eqnarray}\label{quantum-amplification}
(VU)^{k}|+\rangle
&=&\cos(2k+1)\theta |b\rangle +\sin(2k+1)\theta|a\rangle
\nonumber\\
&\simeq& [1-\frac{1}{2}\left((2k+1)\frac{1}{\sqrt{N}}\right)^{2}]|b\rangle
+(2k+1)\frac{1}{\sqrt{N}}|a\rangle.
\end{eqnarray}
The optimal amplification of the state $|a\rangle$ is achieved for 
\begin{eqnarray}
k\sim \frac{\pi}{4}\sqrt{N},
\end{eqnarray}
namely, one can identify the specific state $|a\rangle$ with a probability of the order of unity after the amplification of $k$-number of  steps. The unsorted quantum data is transformed to the data with a probability distribution
\begin{eqnarray}
p_{i}=\langle +|{(VU)^{k}}^{\dagger}|i\rangle\langle i|(VU)^{k}|+\rangle
\end{eqnarray}
that is peaked at $i=a$, which is the location of the  element we are looking for, when  measured by the basis $\{ |i\rangle \}_{i=1}^{N}$. 
 The prediction of the formation of a peak at a specific point $i=a$ is analogous to the prediction of peaks on the screen after the accumulation of data in the double-slit interference experiment. We know the characteristics of the state $|a\rangle$ beforehand, and thus it is essential to locate the position of the state $|a\rangle$ in the set $\{ |i\rangle \}_{i=1}^{N}$.
This is the essence of Grover's algorithm~\cite{grover1,grover2, bennett, zalka, brassard, boyer}.

  It is significant that the above unitary operation is performed on a subspace spanned by a superposition of only two states $\{|a\rangle, |b\rangle\}$ of the entire Hilbert space \eqref{new basis}; if more states are involved, one would need more parameters in addition to $\theta$ to point a given state to the direction of $|a\rangle$. It would be crucial to construct a single well-defined state $|b\rangle$ as a superposition of all the rest of original states using $n$ qubits with $2^{n}=N$, which is strictly orthogonal to other $N-2$ states formed of linear combinations of $\{ |i\rangle \}_{i=1}^{N-1}$, and maintain the strict orthogonality in a realistic quantum computer.
  
  It appears that the entanglement does not play any explicit role in the above analysis. To understand the role of entanglement, one may study the simplest model of two Ising spins  $H=-J{\sigma_{1}}_{z}{\sigma_{2}}_{z}-({\sigma_{1}}_{z}+{\sigma_{2}}_{z})h$ with positive constants $J$ and $h$, for example.  One may start with a product state that has no entanglement
\begin{eqnarray}\label{initial2}
|+\rangle&=&\frac{1}{2}\left(|0\rangle+|1\rangle\right)\otimes\left(|0\rangle+|1\rangle\right) 
\nonumber\\
&=&\frac{1}{2}[|0\rangle\otimes|0\rangle+|0\rangle\otimes|1\rangle+|1\rangle\otimes|0\rangle+|1\rangle\otimes|1\rangle]
\end{eqnarray}
which becomes after marking the lowest energy state $|a\rangle\equiv |0\rangle\otimes|0\rangle$, for example,  by the unitary operator \eqref{detector}
\begin{eqnarray}\label{entanglement}
U|+\rangle
&=&\frac{1}{2}[-|0\rangle\otimes|0\rangle+|0\rangle\otimes|1\rangle+|1\rangle\otimes|0\rangle+|1\rangle\otimes|1\rangle].
\end{eqnarray}
This last expression is no more a product state and in fact entangled, as is confirmed by considering the partial trace of the
density matrix $\rho^{\prime}=U|+\rangle \langle +|U^{\dagger}$ that is reduced to a mixed state~\cite{kwiat, scully}.  In this sense, the entanglement plays a role in Grover's algorithm in the process of identifying the states $|a\rangle$ and $|b\rangle$ in terms of the fundamental qubits.  In fact, any state \eqref{quantum-amplification} is entangled except for the cases with $\sin 2k\theta=0$ for which $(VU)^{k}|+\rangle=\cos\theta |b\rangle +\sin\theta|a\rangle=|+\rangle$ or $\cos(2k+1)\theta=0$ for which $(VU)^{k}|+\rangle=|a\rangle$, if one chooses the initial state as a generalization of \eqref{initial2}.  See also \cite{oh} and references therein.  Incidentally, the initial state such as \eqref{initial2} and the final state $|a\rangle$, which are in principle described by both the classical and quantum formulations, are generally chosen to be product states and not entangled. 

It is also important to recall that the quantum states $|a(t)\rangle$ and 
$|b(t)\rangle$ in \eqref{quantum-amplification} are time dependent in general and contain {\em complex phases},  although the orthonormality, $\langle a(t)|a(t)\rangle=1$, $\langle b(t)|b(t)\rangle=1$ and $\langle a(t)|b(t)\rangle=0$,  are  assumed to be preserved in the present paper.

\section{Classical probability amplification}
To analyze the notion of  classical amplification it is natural to start with a density matrix corresponding to the initial state \eqref{initial-state2} with no quantum coherence
\begin{eqnarray}
\rho=\sum_{i=1}^{N}\frac{1}{N}|i\rangle\langle i|
\end{eqnarray}
with ${\rm Tr}\rho=1$, which gives an equal  probability 
\begin{eqnarray}
p_{i}={\rm Tr}|i\rangle\langle i|\rho=\frac{1}{N}
\end{eqnarray}
for any state $|i\rangle$.  The amplification of the appearance probability of the specific state $|a\rangle$ implies that we realize  a density matrix $\rho=\sum_{i}w_{i}|i\rangle\langle i|$ with $w_{a}\sim 1$ after some physical operation.
We discuss how a classical amplification in the above sense is realized in Grover's algorithm when one assumes that the very rapid dephasing takes place between consecutive Grover's unitary operations.

Starting with a pure state in quantum mechanics 
\begin{eqnarray}
|\psi_{0}\rangle&=&\cos\theta |b\rangle +\sin\theta|a\rangle
\end{eqnarray}
the incoherent mixed state after the complete dephasing, $\overline{|a\rangle\langle b|}=0$ and $\overline{|b\rangle\langle a|}=0$, is defined by 
\begin{eqnarray}
\rho_{0}&=&|\psi_{0}\rangle\langle \psi_{0}|\ \Rightarrow
\overline{\rho}_{0} =
\sin^{2}\theta|a\rangle\langle a| + \cos^{2}\theta |b\rangle\langle b| 
\end{eqnarray}
with
\begin{eqnarray}
{\rm Tr}\overline{\rho}_{0} =
{\rm Tr}\{\sin^{2}\theta|a\rangle\langle a| + \cos^{2}\theta |b\rangle\langle b|\}=1. 
\end{eqnarray}
We emphasize that the idealized dephasing alone does not change the probability of two states $|a\rangle$ and $|b\rangle$. We need some driving force which causes the transition among two states to realize the final states we arrive below; Grover's unitary operation provides precisely this driving force.
  
After the quantum amplification operation \eqref{1.16} of $\overline{\rho}_{0}$,
\begin{eqnarray}
 \rho_{1}&\equiv&VU\overline{\rho}_{0}(VU)^{\dagger}=\sin^{2}\theta VU|a\rangle\langle a|(VU)^{\dagger} + \cos^{2}\theta VU|b\rangle\langle b|(VU)^{\dagger}
\end{eqnarray}
which becomes after the complete dephasing,
\begin{eqnarray}
\overline{\rho}_{1}&=&\overline{VU\overline{\rho}_{0}(VU)^{\dagger}}\nonumber\\
&=&\sin^{2}\theta \overline{VU|a\rangle\langle a|(VU)^{\dagger}} + \cos^{2}\theta \overline{VU|b\rangle\langle b|(VU)^{\dagger}}\nonumber\\
&=&\sin^{2}\theta[\cos^{2}2\theta |a\rangle\langle a|+\sin^{2}2\theta |b\rangle\langle b|]
+\cos^{2}\theta[\cos^{2}2\theta |b\rangle\langle b|+\sin^{2}2\theta |a\rangle\langle a|]\nonumber\\
&=&[\sin^{2}\theta \cos^{2}2\theta +\cos^{2}\theta\sin^{2}2\theta]
|a\rangle\langle a|  \nonumber\\
&+& [\sin^{2}\theta\sin^{2}2\theta + \cos^{2}\theta\cos^{2}2\theta]
|b\rangle\langle b|
\end{eqnarray}
where we used $VU|a\rangle
 =\cos2\theta |a\rangle -\sin2\theta |b\rangle$ and $
VU|b\rangle
 =\cos2\theta |b\rangle +\sin2\theta |a\rangle$ in \eqref{1.16}, 
and thus after the complete decoherence
\begin{eqnarray}
\overline{VU|a\rangle(VU|a\rangle)^{\dagger}}&=&
\overline{(\cos2\theta |a\rangle -\sin2\theta |b\rangle)(\cos2\theta |a\rangle -\sin2\theta |b\rangle)^{\dagger}  }\nonumber\\
&=&\cos^{2}2\theta |a\rangle\langle a|+\sin^{2}2\theta |b\rangle\langle b|, \nonumber\\
\overline{VU|b\rangle(VU|b\rangle)^{\dagger}}&=&
\overline{(\cos2\theta |b\rangle +\sin2\theta |a\rangle )(\cos2\theta |b\rangle +\sin2\theta |a\rangle)^{\dagger}  }\nonumber\\
&=&\cos^{2}2\theta |b\rangle\langle b|+\sin^{2}2\theta |a\rangle\langle a|.
\end{eqnarray}
Note that
\begin{eqnarray}
{\rm Tr}\{\overline{VU|a\rangle(VU|a\rangle)^{\dagger}}\}&=&{\rm Tr}\cos^{2}2\theta |a\rangle\langle a|+{\rm Tr}\sin^{2}2\theta |b\rangle\langle b|=1,\nonumber\\
{\rm Tr}\{\overline{VU|b\rangle(VU|b\rangle)^{\dagger}}\}&=&{\rm Tr}\cos^{2}2\theta |b\rangle\langle b|+{\rm Tr}\sin^{2}2\theta |a\rangle\langle a|=1,
\end{eqnarray}
and thus 
\begin{eqnarray}
{\rm Tr}\overline{\rho}_{1}=1.
\end{eqnarray}
Similarly, we have
\begin{eqnarray}
\overline{\rho}_{2}&=&\overline{VU\overline{\rho}_{1}(VU)^{\dagger}}\nonumber\\
&=&[\sin^{2}\theta (\cos^{4}2\theta +\sin^{4}2\theta) +\cos^{2}\theta(2\sin^{2}2\theta\cos^{2}2\theta)]|a\rangle\langle a| \nonumber\\
&+&[\cos^{2}\theta (\cos^{4}2\theta +\sin^{4}2\theta) +\sin^{2}\theta(2\sin^{2}2\theta\cos^{2}2\theta)]|b\rangle\langle b|
\end{eqnarray}
with
\begin{eqnarray}
{\rm Tr}\overline{\rho}_{2}=1.
\end{eqnarray}
One thus has an iteration formula
\begin{eqnarray}\label{iteration}
\overline{\rho}_{k-1}&=&c_{k-1}|a\rangle\langle a| +d_{k-1}|b\rangle\langle b|,\nonumber\\
 \overline{\rho}_{k}&=&\left( c_{k-1}\cos^{2}2\theta+d_{k-1}\sin^{2}2\theta\right)|a\rangle\langle a| \nonumber\\
 &&+\left( c_{k-1}\sin^{2}2\theta+d_{k-1}\cos^{2}2\theta\right)|b\rangle\langle b|,
 \end{eqnarray}
 with $\overline{\rho}_{0} =\sin^{2}\theta|a\rangle\langle a| + \cos^{2}\theta |b\rangle\langle b|$. This relation is exactly solved as,
\begin{eqnarray}\label{analytic-formula}
\overline{\rho}_{k}&=&c_{k}|a\rangle\langle a| +d_{k}|b\rangle\langle b|,\nonumber\\
c_{k}&=&\cos^{k}4\theta\sin^{2}\theta +\frac{1-\cos^{k}4\theta}{1-\cos4\theta}\sin^{2}2\theta \nonumber\\
&=&\frac{1}{2}-(\frac{1}{2}-\sin^{2}\theta)\cos^{k}4\theta\nonumber\\
&=&\frac{1}{2}-(\frac{1}{2}-\frac{1}{N})[1-\frac{8}{N}(1-\frac{1}{N})]^{k}\nonumber\\
&\simeq&\frac{1}{2}\left( 1-\exp[-8(\frac{k}{N})]\right) 
\end{eqnarray} 
for $N\rightarrow {\rm large}$ with fixed $k/N$~\footnote{The formula \eqref{analytic-formula} is confirmed by mathematical induction. We also note that $[1-\frac{8}{N}(1-\frac{1}{N})]^{k}=\exp[k\ln (1-\frac{8}{N}(1-\frac{1}{N}))]\simeq \exp[-8(\frac{k}{N})]$ for $N\rightarrow {\rm large}$ with fixed $k/N$.}, and $d_{k}=1-c_{k}$. The maximum amplification $c_{k}\simeq 1/2$ is achieved at $k/N=1$ if one considers $k\leq N$; in the classical amplification, \eqref{iteration} shows that $d_{k}$ is amplified if $c_{k}>1/2$. We emphasize that Grover's unitary operation in combination with dephasing is crucial to achieve this final state, which is approximately a thermal equilibrium state for $|a\rangle$ and $|b\rangle$.

More intuitively,  one can confirm that after a $k$-number of  operations
\begin{eqnarray}\label{classical-amplification}
\overline{\rho}_{k} \simeq [\theta^{2}+k(2\theta)^{2}]|a\rangle\langle a|  +[1-\left(\theta^{2}+k(2\theta)^{2}\right)]
|b\rangle\langle b|
\end{eqnarray}
to the accuracy of $O(\theta^{2})=O(\frac{1}{N})$.  This 
$\overline{\rho}_{k}$, in both the exact form \eqref{analytic-formula} and approximate form \eqref{classical-amplification},  satisfies the basic condition of the density matrix
\begin{eqnarray}
{\rm Tr}\overline{\rho}_{k}=1,
\end{eqnarray}
and thus the present way of achieving decoherence is consistent as a {\em model of the very rapid dephasing} of the density matrix between consecutive Grover's unitary operations.

The classical limit of Grover's algorithm induced by a combination of dephasing and Grover's unitary operation thus shows
\begin{eqnarray}\label{classical-probability}
p_{i}={\rm Tr}\{|i\rangle\langle i|\overline{\rho}_{k}\} \simeq (1+4k)/N
\end{eqnarray}
for $i=a$ after $k$-iteration, where we used $\theta^{2}=1/N$ and $p_{i}$ is the probability of finding the state $|i\rangle$. Thus the probability of finding the rest of states becomes $p_{i}={\rm Tr}\{|i\rangle\langle i|\overline{\rho}_{k}\}\simeq 1 - (1+4k)/N$ for $i=b$, or 
$p_{i}={\rm Tr}\{|i\rangle\langle i|\overline{\rho}_{k}\}\simeq \frac{1}{N}(1 - (1+4k)/N)$ for $i\neq a$ if one uses the original states   $\{ |i\rangle \}_{i=1}^{N}$. These formulas agree with the exact result \eqref{analytic-formula} for $N\rightarrow \infty$ with $0 < k/N \ll 1$.   

After the iteration of the order $k\sim N$ times, we thus have the {\em amplified probability} of the order of unity \eqref{classical-probability} (in fact $p_{a} \simeq 1/2$ in the more accurate estimate using \eqref{analytic-formula}) to find the state $a$  by a single trial. The classical amplification defined by our procedure does not change the required total
$k\sim N$ steps to identify  a specific state $a$ although the numerical factor in front of $N$ is generally modified, in contrast to the quantum Grover's algorithm where $k\sim \sqrt{N}$.  The appearance of $k\sim \sqrt{N}$ in Grover's algorithm is due to the fact that we work on the complex probability amplitude in quantum mechanics  instead of the real classical probability, without any irreversible non-unitary effects such as the external thermal agitation.

The reduction of the problem to an effective two-dimensional subspace spanned by $\{|a\rangle, |b\rangle\}$
of the entire Hilbert space \eqref{new basis}, i.e., the target state and an equal  superposition of all the rest of original states,  using the superposition principle of quantum mechanics plays an important role in the present classical amplification also; if more than two states are involved, the natural incoherent density matrix would follow the equal {\em a priori} probability rule for all the involved states (see, for example, \eqref{analytic-formula}) and consequently a smaller amplification of a specific state. 
In other words, we are {\em assuming} that the dephasing is described in terms of the new states defined by the superposition principle \eqref{new basis}  instead of those states described by the original states $\{ |i\rangle \}_{i=1}^{N}$ or the states directly defined by $n$ qubits of the quantum computer with $2^{n}=N$.  

It is also important that we start with an initial state consisting of two states of very unbalanced existence probabilities for the classical amplification.
An illuminating analogy is to imagine a system consisting of two almost degenerate states with very unbalanced occupation numbers $n_{a}\ll n_{b}$ with large $N=n_{a} + n_{b}$; one would then obtain an enormous amplification of $n_{a}$, namely, $n_{a}\sim n_{b}\sim N/2$, by touching the system to a heat bath (a kind of the inverse of the purification by heat bath algorithmic cooling). In this latter case, the driving force to the equilibrium is provided by the thermal agitation, instead of Grover's unitary operation in our analysis.

In connection with the above reduction to an effective two-dimensional subspace of the Hilbert space \eqref{new basis},
an {\em analog} model of Farhi and Gutmann~\cite{farhi}, which does not use entanglement in an obvious way, is interesting. This model uses the notion of time which is missing in the abstract formulation of Grover's algorithm. The time evolution operator in the model is defined by~\cite{farhi} 
\begin{eqnarray}
U(t)&\equiv&e^{-iHt}\nonumber\\
&=&e^{-iEt}\{\cos(xEt) -i\sin(xEt) \left(\begin{array}{cc}
            x&\sqrt{1-x^{2}}\\
            \sqrt{1-x^{2}}&-x
            \end{array}\right)\}
\end{eqnarray}
with the initial condition
\begin{eqnarray}\label{step0}
\psi(0)&=&\left(\begin{array}{c}
            x\\
            \sqrt{1-x^{2}}
            \end{array}\right)\nonumber\\
&=&x |w\rangle +\sqrt{1-x^{2}}|r\rangle,\nonumber\\
\overline{\rho}(0)&=&\overline{\psi(0)\psi^{\dagger}(0)}\nonumber\\
&=&x^{2}|w\rangle\langle w|+(1-x^{2})|r\rangle\langle r|
\end{eqnarray}
where $x$ and $E$ are constant parameters and $|w\rangle$ and $|r\rangle$ correspond to our $|a\rangle$ and $|b\rangle$, respectively.  
The solution of the Schroedinger equation $\psi(t)=U(t)\psi(0)$ is given by 
\begin{eqnarray}
\psi(t)=e^{-iEt}\{\left(x\cos(xEt) -i\sin(xEt)\right) |w\rangle +\sqrt{1-x^{2}}\cos(xEt)|r\rangle \}.
\end{eqnarray}
For an infinitesimal $\Delta t$, which we introduce to realize the situation analogous to Grover's algorithm,
\begin{eqnarray}\label{step1}
U(\Delta t)
&\simeq&e^{-iE\Delta t}\{1-\frac{1}{2}(xE\Delta t)^{2} -i(xE\Delta t) \left(\begin{array}{cc}
            x&\sqrt{1-x^{2}}\\
            \sqrt{1-x^{2}}&-x
            \end{array}\right)\},\nonumber\\
\psi(\Delta t)&\simeq&e^{-i E\Delta t}\{(x -i xE\Delta t) |w\rangle +\sqrt{1-x^{2}}(1-\frac{1}{2}(xE\Delta t)^{2})|r\rangle \},\nonumber\\
\overline{\rho}(\Delta t)&=&\overline{U(\Delta t)\overline{\rho}(0)U^{\dagger}(\Delta t)}\nonumber\\
&\simeq&(x^{2}+(xE\Delta t)^{2})|w\rangle\langle w|+(1-x^{2})(1-(xE\Delta t)^{2})|r\rangle\langle r|,
\end{eqnarray}  
where we used the assumption of complete dephasing between the action of unitary time developments generated by $U(\Delta t)$
\begin{eqnarray}
\overline{U(\Delta t)|w\rangle\langle w|U^{\dagger}(\Delta t)}&=&(1-(xE\Delta t)^{2}(1-x^{2}))|w\rangle\langle w|+
(xE\Delta t)^{2}(1-x^{2})|r\rangle\langle r|,\nonumber\\
\overline{U(\Delta t)|r\rangle\langle r|U^{\dagger}(\Delta t)}&=&(xE\Delta t)^{2}(1-x^{2})|w\rangle\langle w|+
(1-(xE\Delta t)^{2}(1-x^{2}))|r\rangle\langle r|.\nonumber\\
\end{eqnarray}
After $k$-iteration of $U(\Delta t)$ for $xE\Delta t\ll 1$ and choosing $\Delta t$ to be the minimum uncertainty time $E\Delta t=1$ to make the notation simple,  we have
\begin{eqnarray}\label{analog-computer}
\psi(k\Delta t)&\simeq&e^{-ik E\Delta t}\{(x -i kxE\Delta t) |w\rangle +\sqrt{1-x^{2}}(1-\frac{1}{2}(kxE\Delta t)^{2})|r\rangle \}\nonumber\\
&\simeq&e^{-ik}\{x(1 -i k) |w\rangle +\sqrt{1-x^{2}}(1-\frac{1}{2}(kx)^{2})|r\rangle \},\nonumber\\
\overline{\rho}(k\Delta t)&\simeq&(x^{2}+k(xE\Delta t)^{2})|w\rangle\langle w|+(1-x^{2})(1-k(xE\Delta t)^{2})|r\rangle\langle r|
\nonumber\\
&\simeq&x^{2}(1+k)|w\rangle\langle w|+((1-x^{2})-k x^{2})|r\rangle\langle r|.
\end{eqnarray} 
When one sets $x=\theta=1/\sqrt{N}$  in \eqref{analog-computer} following \cite{farhi}, the essence of Grover's formula is recovered. 

The coefficients of the state $|w\rangle$ in \eqref{analog-computer}
show that the transition from the quantum $k\sim \sqrt{N}$ behavior in $\psi(k\Delta t)$, which makes the coefficient of the state $|w\rangle$ to be of the order of unity, to the classical $k\sim N$ behavior in $\overline{\rho}(k\Delta t)$ is caused by the loss of quantum coherence but not by the loss of entanglement which is not defined in the present model. This example indicates that entanglement is crucial in quantum computation to describe the quantum states by a set of qubits as shown in \eqref{entanglement}, but the actual speedup of the quantum computation itself is based on the quantum coherence. Minimum entanglement necessary would rather help speedup the quantum computation itself.  See  \cite{lloyd} for a related comment, and \cite{meyer, bhattacharya, boehm, zhang} for further experimental and theoretical analyses. 
As for conditions other than entanglement, the initial condition with enormously unbalanced existence probabilities for two almost degenerate states in \eqref{step0} is crucial to realize the classical amplification by a combination of dephasing and unitary operation.

\section{Discussion and conclusion}

It is important to understand intuitively why 
it is not possible to search in fewer than $O(\sqrt{N})$ steps in quantum mechanics. In this respect,  it is natural to assume that the classical amplification does not reduce the total search steps to fewer than $O(N)$-steps when formulated in terms of a classical Ising spin, for example. If this is the case, the quantum search process, which deals with the complex probability amplitude rather than the real classical probability itself, can accomplish the search in $O(\sqrt{N})$ steps but not in fewer steps, since otherwise the classical search could be achieved in fewer than $O(N)$ steps. 
To be more explicit, the coherent treatment of the probability amplitude \eqref{quantum-amplification} gives after the $k$-steps of amplification
\begin{eqnarray}
\sim k (2\theta)|a\rangle,
\end{eqnarray} 
while the treatment of the incoherent density matrix \eqref{classical-amplification} gives 
\begin{eqnarray}
\sim k (2\theta)^{2}|a\rangle\langle a|.
\end{eqnarray} 
To achieve the coefficients of the state $|a\rangle$ of the order of unity, one thus needs $k\sim 1/\theta\sim \sqrt{N}$ and $k\sim 1/\theta^{2}\sim N$, respectively.

The present analysis will help understand what the probability amplification is in the classical context, which will simultaneously deepen our understanding of quantum amplification.  The analysis of the effect of dephasing on Grover's algorithm is also crucial in the realistic study of the quantum 
search~\cite{azuma, biham,salas,gawron, botsinis}. Our simple analytic formula \eqref{analytic-formula}, 
\begin{eqnarray}\label{analytic-formula2}
\overline{\rho}_{k}\simeq \frac{1}{2}\left( 1-\exp[-8(\frac{k}{N})]\right)|a\rangle\langle a| +\frac{1}{2}\left( 1+\exp[-8(\frac{k}{N})]\right)|b\rangle\langle b|, 
\end{eqnarray}
will  be useful to understand the dephasing effect in an idealized model. 
 We also mention that other approaches to the classical analogues of Grover's algorithm have been discussed in the past~\cite{lloyd, grover4}. These approaches are, however,  very different from the present analysis that is based on the rapid dephasing of pure quantum states combined with Grover's unitary operation which lead to a classical amplification.

In conclusion, the quantum computation is based on the processing of superposition states~\cite{deutsch}, and Grover's algorithm shows a basic mechanism of the speedups of the quantum search problem, for example, the reduction of classical $N=10^{6}$ steps to $\sqrt{N}=10^{3}$ steps. However, the examination of the practical implementation of Grover's algorithm in  quantum search problems in comparison with the classical computation~\cite{viamontes}
implies that the quantum algorithm is not almighty. For example, a naive application of Grover's algorithm to combinatorial optimization problems such as a search of the ground state in the Ising spin glass models~\cite{barahona} with 150 spins implies $k\sim \sqrt{N}=2^{75}$ which is a very large number. An approximation procedure such as the simulated annealing~\cite{kirkpatrick} is often used in classical digital computers. Practically, an approximation scheme with a clever mixture of quantum and classical computations combining interesting ideas in the literature~\cite{kadowaki, goldstone, cirac} may become relevant. An approximate treatment implies certain decoherence, and it is hoped that the present analysis of dephasing may be useful in such analyses.  
\\
 
One of us (KF) thanks M. Hayashi for a lucid explanation of Grover's algorithm. KF is supported in part by JSPS KAKENHI (Grant No.18K03633).

\end{document}